\documentclass[11pt]{article}
\usepackage{acl-ijcnlp2009}
\usepackage{times}
\usepackage{url}
\usepackage{latexsym}
 \usepackage{graphicx}

\title{Syntax is from Mars while Semantics from Venus!\\ Insights from Spectral Analysis of Distributional Similarity Networks}

\author{Chris Biemann\\
  Microsoft/Powerset, San Francisco\\
  {\tt Chris.Biemann@microsoft.com}  \And
  Monojit Choudhury\\
  Microsoft Research Lab India\\
  {\tt  monojitc@microsoft.com} \AND
  Animesh Mukherjee\\
  Indian Institute of Technology Kharagpur, India\\
  {\tt animeshm@cse.iitkgp.ac.in}
}

\date{}

\begin{document}
\maketitle
\begin{abstract}
We study the global topology of the syntactic and semantic distributional similarity networks for English through the technique of spectral analysis. We observe that while the syntactic network has a hierarchical structure with strong communities and their mixtures, the semantic network has several tightly knit communities along with a large core without any such well-defined community structure.
\end{abstract}

\section{Introduction}
Syntax and semantics are two tightly coupled, yet very different properties of any natural language -- as if one is from ``Mars" and the other from ``Venus". Indeed, this exploratory work shows that the distributional properties of syntax are quite different from those of semantics. {\em Distributional hypothesis} states that the words that occur in the same contexts tend to have similar meanings~\cite{Harris68}. Using this hypothesis, one can define a vector space model for words where every word is a point in some n-dimensional space and the distance between them can be interpreted as the inverse of the semantic or syntactic similarity between their corresponding distributional patterns. Usually, the co-occurrence patterns with respect to the function words are used to define the syntactic context, whereas that with respect to the content words define the semantic context. An alternative, but equally popular, visualization of distributional similarity is through graphs or networks, where each word is represented as nodes and weighted edges indicate the extent of distributional similarity between them. 

What are the commonalities and differences between the syntactic and semantic distributional patterns of the words of a language? This study is an initial attempt to answer this fundamental and intriguing question, whereby we construct the syntactic and semantic distributional similarity network (DSN) and analyze their spectrum to understand their global topology. We observe that there are significant differences between the two networks: the syntactic network has well-defined hierarchical community structure implying a systematic organization of natural classes and their mixtures (e.g., words which are both nouns and verbs); on the other hand, the semantic network has several isolated clusters or the so called {\em tightly knit communities} and a core component that lacks a clear community structure. Spectral analysis also reveals the basis of formation of the natural classes or communities within these networks. These observations collectively point towards a well accepted fact that the semantic space of natural languages has extremely high dimension with no clearly observable subspaces, which makes theorizing and engineering harder compared to its syntactic counterpart.

Spectral analysis is the backbone of several techniques, such as multi-dimensional scaling, principle component analysis and latent semantic analysis, that are commonly used in NLP. In recent times, there have been some work on spectral analysis of linguistic networks as well. Belkin and Goldsmith~\shortcite{Belkin02} applied spectral analysis to understand the struture of morpho-syntactic networks of English words. The current work, on the other hand, is along the lines of Mukherjee et al.~\shortcite{Mukh09}, where the aim is to understand not only the principles of organization, but also the global topology of the network through the study of the spectrum. The most important contribution here, however, lies in the comparison of the topology of the syntactic and semantic DSNs, which, to the best of our knowledge, has not been explored previously.

\section{Network Construction}
The syntactic and semantic DSNs are constructed from a raw text corpus. This work is restricted to the study of English DSNs only\footnote{As shown in~\cite{Nath08}, the basic structure of these networks are insensitive to minor variations in the parameters (e.g., thresholds and number of words) and the choice of distance metric.}.  

{\bf Syntactic DSN}: We define our syntactic network in a similar way as previous works in unsupervised parts-of-speech induction (cf. \cite{Schuet95,Biem06}): The most frequent 200 words in the corpus (July 2008 dump of English Wikipedia) are used as features in a word window of $\pm$2 around the target words. Thus, each target word is described by an 800-dimensional feature vector, containing the number of times we observe one of the most frequent 200 words in the respective positions relative to the target word. In our experiments, we collect data for the most frequent 1000 and 5000 target words, arguing that all syntactic classes should be represented in those. 
A similarity measure between target words is defined by the cosine between the feature vectors. The syntactic graph is formed by inserting the target words as nodes and connecting nodes with edge weights equal to their cosine similarity if this similarity exceeds a threshold $t = 0.66$.

{\bf Semantic DSN}: The construction of this network is inspired by \cite{Lin98}. Specifically, we parsed a dump of English Wikipedia (July 2008) with the XLE parser \cite{Riez02} and extracted the following dependency relations for nouns: Verb-Subject, Verb-Object, Noun-coordination, NN-compound, Adj-Mod. These lexicalized relations act as features for the nouns. Verbs are recorded together with their subcategorization frame, i.e. the same verb lemmas in different subcat frames would be treated as if they were different verbs. We compute log-likelihood significance between features and target nouns (as in~\cite{Dunn93}) and keep only the most significant 200 features per target word. Each feature $f$ gets a feature weight that is inversely proportional to the logarithm of the number of target words it applies on. The similarity of two target nouns is then computed as the sum of the feature weights they share. For our analysis, we restrict the graph to the most frequent 5000 target common nouns and keep only the 200 highest weighted edges per target noun. Note that the degree of a node can still be larger than 200 if this node is contained in many 200 highest weighted edges of other target nouns. 

\section{Spectrum of DSNs}

\begin{figure}
\centering
\includegraphics[width=2in]{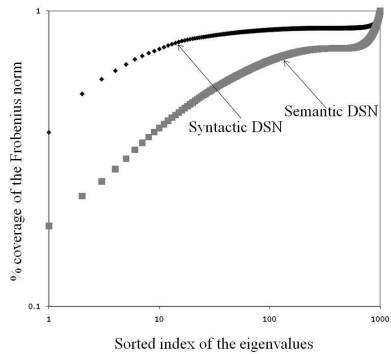}
\caption{The spectrum of the syntactic and semantic DSNs of 1000 nodes.}
\label{fig:spectrum}
\end{figure}

{\em Spectral analysis} refers to the systematic study of the eigenvalues and eigenvectors of a network. Although here we study the spectrum of the adjacency matrix of the weighted networks, it is also quite common to study the spectrum of the Laplacian of the adjacency matrix (see for example, Belkin and Goldsmith (2002)). Fig.~\ref{fig:spectrum} compares the spectrum of the syntactic and semantic DSNs with 1000 nodes, which has been computed as follows. First, the 1000 eigenvalues of the adjacency matrix are sorted in descending order. Then we compute the spectral coverage till the $i$th eigenvalue by adding the squares of the first $i$ eigenvalues and normalizing it by the sum of the squares of all the eigenvalues - a quantity also known as the Frobenius norm of the matrix. 

We observe that for the semantic DSN the first 10 eigenvalues cover only 40\% of the spectrum and the first 500 together make up 75\% of the spectrum. On the other hand, for the syntactic DSN, the first 10 eigenvalues cover 75\% of the spectrum while the first 20 covers 80\%. In other words, the structure of the syntactic DSN is governed by a few (order of 10) significant principles, whereas that of the semantic DSN is controlled by a large number of equally insignificant factors. 

The aforementioned observation has the following alternative, but equivalent interpretations: (a) the syntactic DSN can be clustered in lower dimensions (e.g., 10 or 20) because, most of the rows in the matrix can be approximately expressed as a linear combination of the top 10 to 20 eigenvectors. Furthermore, the graceful decay of the eigenvalues of the syntactic DSN implies the existence of a hierarchical community structure, which has been independently verified by Nath et al.~\shortcite{Nath08} through analysis of the degree distribution of such networks; and (b) a random walk conducted on the semantic DSN will have a high tendency to  drift away very soon from the semantic class of the starting node, whereas in the syntactic DSN, the random walk is expected to stay within the same syntactic class for a long time. Therefore, it is reasonable to advocate that characterization and processing of syntatic classes is far less confusing than that of the semantic classes -- a fact that requires no emphasis.

\begin{figure}
\centering
\includegraphics[width=2.5in]{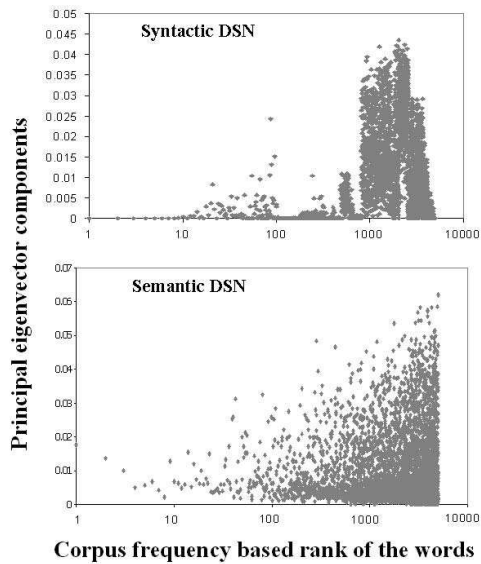}
\caption{Plot of corpus frequency based rank vs. eigenvector centrality of the words in the DSNs of 5000 nodes.}
\label{fig:evc}
\end{figure}

\section{Eigenvector Analysis}

The first eigenvalue tells us to what extent the rows of the adjacency matrix are correlated and therefore, the corresponding eigenvector is not a dimension pointing to any classificatory basis of the words. However, as we shall see shortly, the other eigenvectors corresponding to the significantly high eigenvalues are important classificatory dimensions.  

Fig~\ref{fig:evc} shows the plot of the first eigenvector component (aka {\em eigenvector centrality}) of a word versus its rank based on the corpus frequency. We observe that the very high frequency (i.e., low rank) nodes in both the networks have low eigenvector centrality, whereas the medium frequency nodes display a wide range of centrality values. However, the most striking difference between the networks is that while in the syntactic DSN the centrality values are approximately normally distributed for the medium frequency words, the least frequent words enjoy the highest centrality for the semantic DSN. Furthermore, we observe that the most central nodes in the semantic DSN correspond to semantically unambiguous words of similar nature (e.g., deterioration, abandonment, fragmentation, turmoil). This indicates the existence of several ``tightly knit communities consisting of not so high frequency words" which pull in a significant fraction of the overall centrality. Since the high frequency words are usually polysemous, they on the other hand form a large, but non-cliqueish structure at the core of the network with a few connections to the tightly knit communities. This is known as the tightly knit community effect (TKC effect) that renders very low centrality values to the ``truly" central nodes of the network~\cite{Lempel00}. The structure of the syntactic DSN, however, is not governed by the TKC effect to such an extreme extent. Hence, one can expect to easily identify the natural classes of the syntactic DSN, but not its semantic counterpart.

\begin{figure}
\centering
\includegraphics[width=2in]{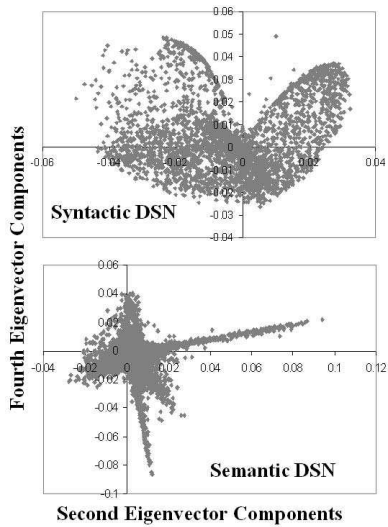}
\caption{Plot of the second vs. fourth eigenvector components of the words in the DSNs.}
\label{fig:highev}
\end{figure}

In fact, this observation is further corroborated by the higher eigenvectors. Fig.~\ref{fig:highev} shows the plot of the second eigenvector component versus the fourth one for the two DSNs consisting of 5000 words. It is observed that for the syntactic network, the words get neatly clustered into two sets comprised of words with the positive and negative second eigenvector components. The same plot for the semantic DSN shows that a large number of words have both the components close to zero and only a few words stand out on one side of the axes -- those with positive second eigenvector component and those with negative fourth eigenvector component. In essence, none of these eigenvectors can neatly classify the words into two sets -- a trend which is observed for all the higher eigenvectors (we conducted experiments for up to the twentieth eigenvector).

Study of the individual eignevectors further reveals that the nodes with either the extreme positive or the extreme negative components have strong linguistic correlates. For instance, in the syntactic DSN, the two ends of the second eigenvector correspond to nouns and adjectives; one of the ends of the fourth, fifth, sixth and the twelfth eigenvectors respectively correspond to location nouns, prepositions, first names and initials, and verbs. In the semantic DSN, one of the ends of the second, third, fourth and tenth eigenvectors respectively correspond to professions, abstract terms, food items and body parts. One would expect that the higher eigenvectors (say the 50$^{\textrm{th}}$ one) would show no clear classificatory basis for the syntactic DSN, while for the semantic DSN those could be still associated with prominent linguistic correlates.

\section{Conclusion and Future Work}
Here, we presented some initial investigations into the nature of the syntactic and semantic DSNs through the method of spectral analysis, whereby we could observe that the global topology of the two networks are significantly different in terms of the organization of their natural classes. While the syntactic DSN seems to exhibit a hierarchical structure with a few strong natural classes and their mixtures, the semantic DSN is composed of several tightly knit small communities along with a large core consisting of very many smaller ill-defined and ambiguous sets of words. To visualize, one could draw an analogy of the syntactic and semantic DSNs respectively to ``crystalline" and ``amorphous" solids. 

This work can be furthered in several directions, such as, (a) testing the robustness of the findings across languages, different network construction policies, and corpora of different sizes and from various domains; (b) clustering of the words on the basis of eigenvector components and using them in NLP applications such as unsupervised POS tagging and WSD; and (c) spectral analysis of WordNet and other manually constructed ontologies.

\section*{Acknowledgement}
CB and AM are grateful to Microsoft Research India, respectively for hosting him while this research was conducted, and financial support.

\end{document}